\PassOptionsToPackage{bookmarks={false}}{hyperref}
\documentclass[preprint]{vgtc}               

\ifpdf
  \pdfoutput=1\relax                   
  \usepackage{graphicx}                
  \DeclareGraphicsExtensions{.pdf,.png,.jpg,.jpeg} 
\else
  \usepackage{graphicx}                
  \DeclareGraphicsExtensions{.eps}     
\fi%

\graphicspath{{figures/}{pictures/}{images/}{./}} 

\usepackage{microtype}                 
\PassOptionsToPackage{warn}{textcomp}  
\usepackage{textcomp}                  
\usepackage{mathptmx}                  
\usepackage{times}                     
\usepackage{cite}                      
\usepackage{tabu}                      
\usepackage{booktabs}                  
\usepackage{paralist}
\usepackage{amsmath}
\usepackage{multirow}
\usepackage[table,xcdraw]{xcolor}
\usepackage{flushend}

\onlineid{1037}

\vgtccategory{Research}

\vgtcinsertpkg

\newcommand{\rev}[1]{{\color{black} #1}}

\title{Improving Engagement of Animated Visualization with \\ Visual Foreshadowing}

\author{Wenchao Li\thanks{e-mail: wlibs@connect.ust.hk}\\ %
        \parbox{1.4in}{\scriptsize \centering The Hong Kong University of Science and Technology} %
\and Yun Wang\thanks{e-mail: wangyun@microsoft.com}\\ %
    \parbox{1.4in}{\scriptsize \centering Microsoft Research Asia} %
\and Haidong Zhang\thanks{e-mail: haidong.zhang@microsoft.com}\\ %
     \parbox{1.4in}{\scriptsize \centering Microsoft Research Asia} %
\and Huamin Qu\thanks{e-mail: huamin@cse.ust.hk}\\ 
     \parbox{1.4in}{\scriptsize \centering The Hong Kong University of Science and Technology}} %

\abstract{
Animated visualization is becoming increasingly popular as a compelling way to illustrate changes in time series data. However, maintaining the viewer’s focus throughout the entire animation is difficult because of its time-consuming nature. Viewers are likely to become bored and distracted during the ever-changing animated visualization. Informed by the role of foreshadowing that builds the expectation in film and literature, we introduce \textit{visual foreshadowing} to improve the engagement of animated visualizations. In specific, \rev{we propose designs of visual foreshadowing that engage the audience while watching the animation.} To demonstrate our approach, we built a proof-of-concept animated visualization authoring tool that incorporates visual foreshadowing techniques with various styles. Our user study indicates the effectiveness of our foreshadowing techniques on improving engagement for animated visualization. %
} 

\CCScatlist{
  \CCScatTwelve{Human-centered computing}{Visu\-al\-iza\-tion}{Visu\-al\-iza\-tion design and evaluation methods}{};
  \CCScatTwelve{Computing methodologies}{Com\-pu\-ter graphics}{\rev{Animation}}{}
}

\begin{document}


\maketitle

\section{Introduction} 
In recent years, animated visualization has been widely adopted to show changes in time series data, because of its inherent nature of presenting temporal evolution over time~\cite{chevalier2016animations,robertson2008effectiveness}. 
\rev{When used appropriately, animated visualization can be attractive and effective to convey information~\cite{tversky2002animation, heer2007animated}.}
A well-known example is Hans Rosling's Gapminder presentation using the animated bubble chart to review the financial and physical well-being changes of different countries over 200 years~\cite{rosling2009gapminder}. 
This work has garnered more than 8 million views from all over the world~\cite{kosara2013storytelling} and has drawn considerable public attention toward data visualization. 
Since then, more and more animated visualizations have emerged to present changes in data. 
For example, \emph{Top 15 Best Global Brands Ranking}~\cite{top15best} and \emph{The History of the World's Best Go Players}~\cite{thehistoryof} use an animated bar chart to show changes in rankings. 

However, without additional guidance from a presenter, general audience might feel difficult to pay attention to key information and reluctant to consume an entire animation. Although animated visualizations are visually appealing, people are not sure where to focus when multiple changes simultaneously occur. Moreover, the ever-changing nature of animated visualization makes it hard for people to retain their attention for a long time. 

In film and literature, foreshadowing is widely adopted as an important narrative tactic to engage viewers and readers~\cite{chatman1980story}. This narrative tactic uses implications early in a story to indicate the subsequent emergence of a relevant occurrence in the plot~\cite{wang2016animated,moura2018foreshadowing}. 
Therefore, we propose to adopt {\em visual foreshadowing\/} for data storytelling to improve the engagement of animated visualization. 
\rev{Visual foreshadowing guides audience's attention to significant changes in data and raise expectations for forthcoming events.

In this study, we formulate and apply visual foreshadowing on animated bar chart, which is a common animated visualization for ranking changes.} 
\rev{We demonstrate four examples of visual foreshadowing that can be categorized into two major types: \emph{explicit} foreshadowing (that openly suggests the outcome) and \emph{implicit} foreshadowing (that leaves subtle clues by hinting the relevant items). 
To learn the efficacy of engagement enhancement of our approach, we implement a proof-of-concept system that generates animated bar charts with diverse visual foreshadowing techniques. With this new animated visualization authoring tool, we support the easy creation of visual foreshadowing effects for specific item(s). Our system facilitates adding, previewing, and editing visual foreshadowing for the selected items with different temporal ranking data. 
Our user study results suggest that the proposed visual foreshadowing techniques are useful to engage the audience of animated visualization. 
} 

\section{Related Work} 
A large variety of animations being applied to different domains. Thus, the need for emphasizing crucial parts of animation has motivated researchers to study attention guidance for dynamic visualization~\cite{chevalier2014not,du2015trajectory, wang2017vector,lu2020illustrating}. For example, in the education domain, De Koning et al.~\cite{de2009towards} attempted to transfer successful cueing approaches from static visualization to animation for instructional design. The authors proposed a framework that classifies three cueing functions, selection, organization, and integration. This work suggested developing visual cues for animation instead of borrowing the effective ones for static representations. 
The more recent work of De Koning et al.~\cite{de2017attention}, summarized visual cues to direct learner attention to key information, including basic arrows, colored circles, and other visual cues embedded within graphical entities of dynamic visualization. \rev{
In an effort to locate task-relevant information, Etemadpour et al.~\cite{etemadpour2017density} and Chen et al. ~\cite{chen2018using} applied motions to data points, and De Koning et al.~\cite{de2011attention} studied the role of presentation speed in attention cueing. In dynamic narrative visualization, Waldner et al.~\cite{waldner2014attractive} studied the flicker effect and found a good trade-off between attraction effectiveness and the subjective annoyance caused by flickering. Many of the considerations from these works informed the design of our visual foreshadowing. 
}

\rev{
Despite the availability of existing studies, the topic of how visual cues help improve engagement in animation or dynamic visualization is still under exploration. The work by Wang et al.~\cite{wang2016animated} is the most relevant to our approach. The authors first adopted the term ``foreshadowing'' to help enhance the narratives of clickstream data visualization. However, their techniques are employed in the timeline, and the visual effects are limited to animated stacked graph. In the current research, we formally define visual foreshadowing and propose design examples for animated visualization. 
}


\section{Visual Foreshadowing}
Foreshadowing is a narrative tactic that is widely used in film and literature. It often appears in the early stages of a plot because it can subtlety create tension and set viewers' expectations regarding how the story will unfold. We conduct an extensive literature review of the foreshadowing techniques used in the literature, film, and video games domains\rev{~\cite{kong2018frames,bordwell1985narration,martin2002writer,bae2013modeling,turner1994creative,aarseth2012narrative}, analyze the existing visual cues in data visualization~\cite{waldner2014attractive, wang2016animated, kong2017internal}, and formally define visual foreshadowing. Taking bar chart race as an example, we initialize visual foreshadowing designs for animated visualization.}

\subsection{Definition}
We see visual foreshadowing as a new form of animation effect that appears before the critical events during the playback of animated visualization to set the viewer's expectations.
Basing from previous research on animation~\cite{ge2020canis}, we define visual foreshadowing for animated visualization from three perspectives, namely, \textit{visual effect}, \textit{timing}, and \textit{duration}. Accordingly, visual foreshadowing can be formalized as a 3-tuples: 
\begin{displaymath}
\textit{Visual Foreshadowing} := (\textit{visual effect(s)}, \textit{timing(s)},  \textit{\textbf{duration}(s)}).
\end{displaymath}

Visual foreshadowing can have one or multiple \textit{visual effects}.
The visual effects in visual foreshadowing can be a textual statement or the ones that are attached to specific visual elements. Under the framework,  different visual cues (e.g., flickering or pointing with arrows) can be easily introduced to extend the foreshadowing effects. 

\rev{Foreshadowing can be achieved implicitly or explicitly~\cite{bae2008use, wang2016animated}. Following this taxonomy, the \textit{visual effects} adopted in visual foreshadowing should also have two categories. \textit{Explicit foreshadowing} occurs when an outcome is directly indicated, which explicitly gives viewers a piece of information to entice them to want more. 
Therefore, the audience is directly informed of the key information or the final outcome. \textit{Implicit foreshadowing} appears when an outcome is indirectly hinted, which leaves subtle clues about a future event by suggesting related items. 
The upcoming event is only apparent to viewers after it has occurred. On the basis of prior work on visual cue preference and highlighting interventions for static visualizations~\cite{carenini2014highlighting}, the visual effects for indicating relevant visual elements in indirect foreshadowing can include adding a contour and setting the transparency. By using the visual cues, the author can draw viewers' attention to the target items.}

Given the specified visual effect, \textit{timing} is used to specify when the visual effects start, while \textit{duration} is to control the time length of the visual effects. Timing and duration should be chosen for each foreshadowing visual effect.
The author can define when to start building the viewer's expectations and when to stop before the crucial event occurs. 

\begin{figure}[tb]
 \centering 
 \includegraphics[width=0.95\columnwidth]{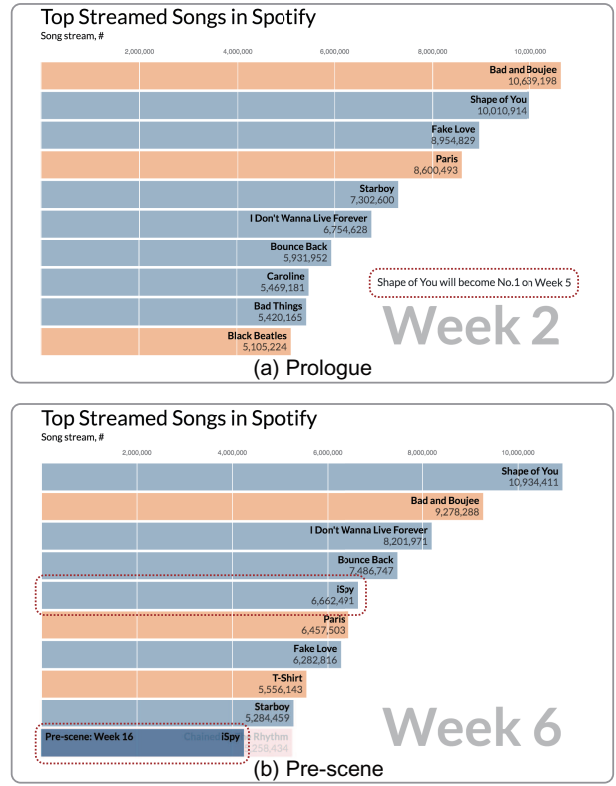}
 \caption{The visual effects of two explicit foreshadowing before the key event occurs. The Prologue effect is designed to build the expectation literally, while the Pre-scene effect is designed to achieve it visually.  \rev{Data are collected from Spotify (https://spotifycharts.com/regional), which provides the historical popularity rankings of the songs streamed by the users. The colors of the bars encode the type of the singers.}}
 \label{fig:explicit}
 \vspace{-4mm}
\end{figure}

\subsection{Example Foreshadowing Effects}
\rev{The design of foreshadowing can be numerous. Considering the design space of possible visual cues~\cite{kong2017internal}, we strategically select the common cueing methods to ease the burden of understanding the visual designs. 
We demonstrate two explicit and two implicit foreshadowing techniques to enhance animated visualization with two different datasets. }

\begin{figure}[tb]
 \centering 
 \includegraphics[width=0.95\columnwidth]{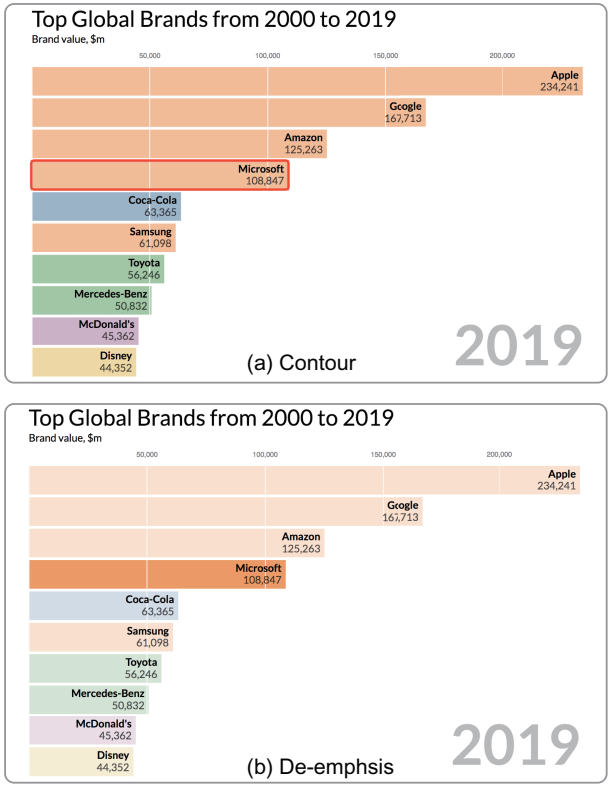}
 \caption{The visual effects of two implicit foreshadowing. Contour and De-emphasis effects are used to highlight the interesting bar to attract the viewer's attention. \rev{Data are collected from Interbrand (https://www.interbrand.com/best-brands), which provides the historical rankings of the world’s leading brands. The colors of the bars encode the category of the brands.}}
 \label{fig:implicit}
 \vspace{-4mm}
\end{figure}

\begin{compactitem}
\item \textit{\textbf{Prologue.}} 
\rev{Prologue is an explicit foreshadowing example. Inspired by the study of Kong et al.~\cite{kong2018frames} that the perceived main message of a visualization can be influenced by the slant of the title}, the foreshadowing design is to insert additional text description to suggest the forthcoming events (see \autoref{fig:explicit}(a)). After noticing the caption, the audience's focus is guided to the target bar so that they can see how the bar animates to meet their expectation. 
\item \textit{\textbf{Pre-scene.}} Pre-scene is an explicit foreshadowing example. The goal of the design is to arouse the viewer's interest to see how the selected bar would change to the final state. This direct foreshadowing design builds the expectation visually, where the final ranking and the length of the bar are directly visualized. \rev{As illustrated in \autoref{fig:explicit}(b), the final state of the \textit{iSpy} item is directly shown in the animated visualization. People may be curious about how the suggested item of the current state turns out.}
\item \textit{\textbf{Contour.}} 
\rev{Contour is an implicit foreshadowing example.
The visual effect of the foreshadowing technique is drawing a contour around the target bar to attract the audience's attention before important change occurs. In contrast to the explicit foreshadowing techniques, this type of visual foreshadowing only highlights the relevant items without disclosing what will happen in the future. As shown in \autoref{fig:implicit}(a), the bar with contour is suggesting some interesting change will happen to it. }
\item \textit{\textbf{De-emphasis.}} De-emphasis is an implicit foreshadowing example. Similar to the Contour effect, interesting items will be highlighted before key events occur. As shown in \autoref{fig:implicit}(b), the De-emphasis effect is implemented by setting transparency. The relevant items hold the original transparency, whereas the transparency of those irrelevant bars is set to 20\%. 

\end{compactitem}

\section{Visual Foreshadowing Authoring}
\begin{figure*}[tb]
 \centering 
 \includegraphics[width=0.95\linewidth]{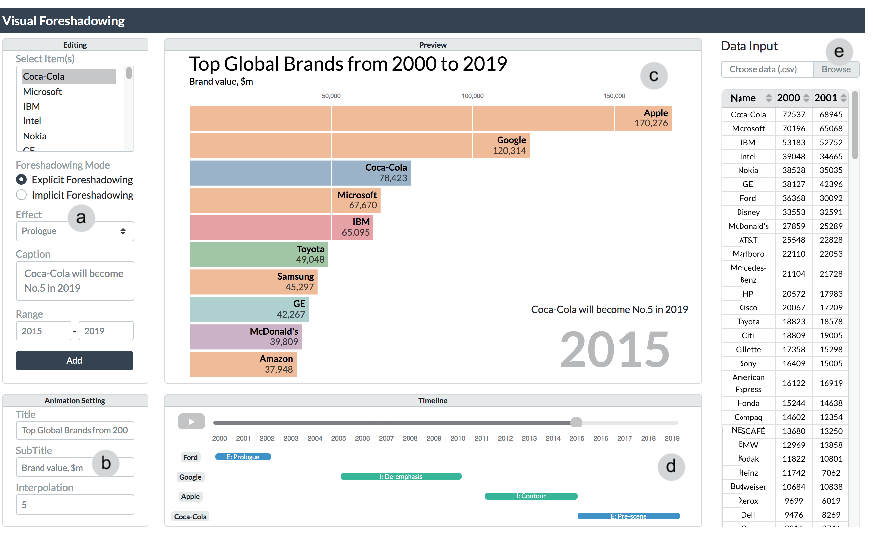}
 \caption{Our prototype system for visual foreshadowing authoring. The system mainly consists of (a) editing widgets for visual foreshadowing, (b) basic settings for animated visualization, (c) animated visualization preview, (d) timeline for visual foreshadowing, and (e) data view. The author can add visual foreshadowing for a specific item and get an overall review of the timeline. }
 \label{fig:poc}
 \vspace{-2mm}
\end{figure*}

To evaluate the visual foreshadowing design, we built a proof-of-concept system for animated visualization authoring. The user interface of the prototyping system consists of a foreshadowing configuration panel (\autoref{fig:poc}(a)), a basic animation setting widget (\autoref{fig:poc}(b)), an animation preview panel (\autoref{fig:poc}(c)), a timeline panel (\autoref{fig:poc}(d)), and a data loading panel (\autoref{fig:poc}(e)). 
We provide additional details about the system with an example scenario for the creation of an animated bar chart. 

Before applying the visual foreshadowing to the animated bar charts, the user needs to load the temporal data of ranking changes. The underlying data can be updated by editing the data table. 
As shown in \autoref{fig:poc}(e), each row of the data table represents a ranking item, while the columns show the temporal dimension. 

After loading the CSV data file of the chart, an initial static bar chart will be displayed in the preview panel (\autoref{fig:poc}(c)), and the animated visualization  illustrating the ranking changes can be played. 
To apply the visual foreshadowing to the animated bar chart, the user need to select the specific items and create the associated visual foreshadowing effect in the foreshadowing editing widget (\autoref{fig:poc}(a)). 
In our current proof-of-concept implementation, we provide the control of visual foreshadowing in terms of foreshadowing type, visual effects, and time range. 

For instance, a user wants to add an explicit foreshadowing effect to suggest the rank of the Coca-Cola company in 2019. The user can obtain the corresponding bar by choosing the item in the item list. Then, the user needs to specify the settings for explicit foreshadowing by selecting the effect type, providing relevant words, and choosing the time range to show the visual effects. Finally, the resulting foreshadowing effect with time range is indicated in the timeline widget (\autoref{fig:poc}(d)). The user can preview the synthesized animated visualization with additional visual foreshadowing effects. 

\section{Evaluation}
Our approach is designed to build the expectation and guide the audience's attention to watch the animated visualization in an engaging manner. 
Our preliminary observations show that the viewers are often unaware where to focus without a narrator  presenting, explaining, and highlighting the key messages when frequent changes occur in the ranking data. 
Through introducing timely indicators, we provide additional clues for the animated visualization of temporal data. 
To understand if our visual foreshadowing techniques could enhance the engagement of animated visualization, we conducted a user study to compare the effects of playing the ranking animation with and without visual foreshadowing effects.

\subsection{Study Design}
We recruited 12 participants (5 males and 7 females; aged 20-27 [median = 23.3, SD = 2.5]), denoted as P1-P12, for this study. 
The participants were mostly graduate and undergraduate students in a local university and with different backgrounds (i.e., computer science, life science, and arts). 
We started our user study by introducing the dataset background and the encoding scheme of the proposed visualization. 
\rev{We used two different ranking change visualizations of global top brands with two different time periods and provided the participants with two visual foreshadowing settings (i.e., with and without) for comparison. Different types of visual foreshadowing designs are covered under the condition with foreshadowing.} Then, we asked each participant to complete a survey about the overall user engagement of the animated visualization.
The survey consisted of 12 subjective questions referenced from previous studies~\cite{amini2018hooked}. Each question was rated on a 7-point Likert scale (1 = strongly disagree, 7 = strongly agree).
In addition, we introduced the proof-of-concept system to each participant and encouraged them to freely explore the system to generate visual foreshadowing for the ranking animations. 
Finally, the user study ended with a semi-structured interview for collecting feedback from each participant. Each participant took approximately 20 minutes to complete the whole study, including the informal interview. 

\subsection{Results}
\autoref{table:survey} shows the results of three categories of engagement measures. 
Overall, the animated visualization with visual foreshadowing achieves higher subjective ratings for the one without visual foreshadowing, indicating the effectiveness of our visual foreshadowing design on engagement improvement. 
In the interview session, almost all the participants valued the visual foreshadowing techniques. 
One participant (P3) commented that ``{\em The additional visual effects are useful and make the animation more like a story. Otherwise, I don't know where to look at and forget almost all the changes.\/}'' Another participant (P12) said ``{\em I built the anticipation when I saw the upcoming icons on the timeline, which sets me a clearer goal for where to concentrate on.\/}''

However, some participants also offered suggestions for the foreshadowing design. One of the participants (P10) said, ``{\em I think it would be very interesting to include the slow-down and zooming effects to draw our attentions.\/}'' \rev{A participant (P8) pointed out that a limited number of visual foreshadowing can improve the overall engagement of the animation. However, adding excessive visual foreshadowing would be overwhelming to the audience. Moreover, when visual foreshadowing is introduced, the audience expected it to be placed long enough before the event occurs. This setting will give them more joy when they come back through the data story and find the message left before. }

\begin{table}[tb]
\caption{Average ratings of the user engagement survey questions  \newline (1 = strongly disagree, 7 = strongly agree).}
\centering
\begin{tabular}{cccc}
\toprule[1pt]
\textbf{Assessment}                     & \textbf{Foreshadowing}         & \textbf{Mean}                & \textbf{SD}                  \\
\hline
                                        & Without                           & 4.13                         & 0.61                         \\
\multirow{-2}{*}{Enjoyment}              & \cellcolor[HTML]{EFEFEF}With & \cellcolor[HTML]{EFEFEF}6.54 & \cellcolor[HTML]{EFEFEF}0.54 \\
\hline
                                        & Without                           & 3.71                         & 1.03                         \\

\multirow{-2}{*}{Focused Attention}     & \cellcolor[HTML]{EFEFEF}With & \cellcolor[HTML]{EFEFEF}6.17 & \cellcolor[HTML]{EFEFEF}0.72 \\
\hline
                                        & Without                           & 4.50                         & 1.04                         \\

\multirow{-2}{*}{Cognitive Involvement} & \cellcolor[HTML]{EFEFEF}With & \cellcolor[HTML]{EFEFEF}6.04 & \cellcolor[HTML]{EFEFEF}0.58 \\
\hline
\end{tabular}
\vspace{2mm}

\label{table:survey}
\vspace{-7mm}
\end{table}

\section{Discussion}
\paragraph{Future Extensions of Visual Foreshadowing Designs} The goal of the visual foreshadowing in our work is to entice the viewer to consume the entire animation instead of the one with tedious changes. 
As a first step, we propose four visual foreshadowing design examples focus on the animated bar chart of temporal ranking data. However, visual foreshadowing should not be limited to specific charts types. For example, more visual effects and even dynamic visual cues can be introduced and combined with the temporal aspect of our visual foreshadowing. Moreover, other forms of animated visualizations can be adopted. For example, the foreshadowing techniques of direct text description and indirect item indication can be applied to animated line charts or animated scatterplots without too much modification. The visual effects, given their timing and duration settings, can still be applied to these common animated visualizations.  

\paragraph{In-depth Investigation on Effectiveness}
In our user study, we investigated visual foreshadowing with several visual effects generally. \rev{Interestingly, participants had a stronger preference on the foreshadowing designs that only leaves subtle clues rather than the ones that openly suggest the events was empirically observed. For instance, compared with participants that tested with the Pre-scene effect, participants that tested with the De-emphasis effect were more willing to wait until the end of an animated transition to obtain the final outcome. However, which visual foreshadowing design and which foreshadowing setting (e.g., timing, and duration) are the most effective ones remain unclear. What about the combinations of visual foreshadowing for engagement improvement? In addition, we studied general engagement improvement on a rather small tested data. Whether or not viewers will behave differently in more complicated data is also interesting to study. Examining these ideas requires further investigation, which is outside the scope of this work. We encourage follow-up studies to explore the different aspects of visual foreshadowing used in animated visualization.
}
\section{Conclusion}
In this work, we introduced visual foreshadowing, a design that enhances animated visualization with effective visual cues ahead of time. The visual cues with proper timing prepare the audience with critical patterns of temporal data and enhance temporal change recognition. We build a proof-of-concept system to support foreshadowing authoring and further investigation. We demonstrate our visual foreshadowing approach with a qualitative evaluation. Results show that our method can guide participants' preparation for the upcoming events in the generated animated ranking visualization. The subjective preference ratings reveal that animated visualization with visual foreshadowing are engaging to show temporal data changes. 
In the future, we plan to investigate other design choices (e.g., the design of visual cues) with different levels of data complexities. We will also integrate visual foreshadowing into other forms of animated visualization, such as dynamic line chart and scatter plot, to provide effective animated narrative visualizations.

\acknowledgments{
The authors thank the anonymous reviewers for their valuable comments. This work was supported in part by a grant from Microsoft Research Asia.
}

\bibliographystyle{abbrv-doi}

\bibliography{reference}
\end{document}